\documentclass[twocolumn,showpacs,aps,prl,superscriptaddress]{revtex4}

\usepackage{graphicx}
\usepackage{dcolumn}
\usepackage{amsmath}
\usepackage{epsfig}
\RequirePackage{xspace}

\usepackage{relsize}
\def\babar{\mbox{\slshape B\kern-0.1em{\smaller A}\kern-0.1em
    B\kern-0.1em{\smaller A\kern-0.2em R}}}

\def\g     {\ensuremath{\gamma}\xspace}
\def\q     {\ensuremath{q}\xspace}

\def\qqbar {\ensuremath{q\overline q}\xspace}
\def\u     {\ensuremath{u}\xspace}
\def\ubar  {\ensuremath{\overline u}\xspace}

\def\d     {\ensuremath{d}\xspace}

\def\s     {\ensuremath{s}\xspace}

\def\c     {\ensuremath{c}\xspace}
\def\cbar  {\ensuremath{\overline c}\xspace}

\def\b     {\ensuremath{b}\xspace}

\def\piz   {\ensuremath{\pi^0}\xspace}

\def\pip   {\ensuremath{\pi^+}\xspace}
\def\pim   {\ensuremath{\pi^-}\xspace}

\def\Kbar  {\kern 0.2em\overline{\kern -0.2em K}{}\xspace}

\def\Kz    {\ensuremath{K^0}\xspace}
\def\Kzb   {\ensuremath{\Kbar^0}\xspace}
\def\KzKzb {\ensuremath{\Kz \kern -0.16em \Kzb}\xspace}
\def\Kp    {\ensuremath{K^+}\xspace}
\def\Km    {\ensuremath{K^-}\xspace}

\def\Kmp   {\ensuremath{K^\mp}\xspace}
\def\KpKm  {\ensuremath{\Kp \kern -0.16em \Km}\xspace}
\def\KS    {\ensuremath{K^0_{\scriptscriptstyle S}}\xspace}

\def\Kstar   {\ensuremath{K^*}\xspace}

\def\Dbar    {\kern 0.2em\overline{\kern -0.2em D}{}\xspace}
\def\Db      {\ensuremath{\Dbar}\xspace}
\def\Dz      {\ensuremath{D^0}\xspace}
\def\Dzb     {\ensuremath{\Dbar^0}\xspace}
\def\DzDzb   {\ensuremath{\Dz {\kern -0.16em \Dzb}}\xspace}
\def\Dp      {\ensuremath{D^+}\xspace}
\def\Dm      {\ensuremath{D^-}\xspace}

\def\DpDm    {\ensuremath{\Dp {\kern -0.16em \Dm}}\xspace}
\def\Dstar   {\ensuremath{D^*}\xspace}

\def\Dstarz  {\ensuremath{D^{*0}}\xspace}

\def\Dstarp  {\ensuremath{D^{*+}}\xspace}

\def\B       {\ensuremath{B}\xspace}
\def\Bbar    {\kern 0.18em\overline{\kern -0.18em B}{}\xspace}

\def\BB      {\ensuremath{B\Bbar}\xspace} 
\def\Bz      {\ensuremath{B^0}\xspace}
\def\Bzb     {\ensuremath{\Bbar^0}\xspace}
\def\BzBzb   {\ensuremath{\Bz {\kern -0.16em \Bzb}}\xspace}
\def\Bu      {\ensuremath{B^+}\xspace}
\def\Bub     {\ensuremath{B^-}\xspace}
\def\Bp      {\ensuremath{\Bu}\xspace}
\def\Bm      {\ensuremath{\Bub}\xspace}

\def\Bmp     {\ensuremath{B^\mp}\xspace}
\def\BpBm    {\ensuremath{\Bu {\kern -0.16em \Bub}}\xspace}

\def\BorBbar    {\kern 0.18em\optbar{\kern -0.18em B}{}\xspace}
\def\DorDbar    {\kern 0.18em\optbar{\kern -0.18em D}{}\xspace}
\def\KorKbar    {\kern 0.18em\optbar{\kern -0.18em K}{}\xspace}

\mathchardef\Upsilon="7107
\def\Y#1S{\ensuremath{\Upsilon{(#1S)}}\xspace}% no space before {...}!

\def\FourS {\Y4S}

\mathchardef\Deltares="7101
\mathchardef\Xi="7104
\mathchardef\Lambda="7103
\mathchardef\Sigma="7106
\mathchardef\Omega="710A

\def\Deltabar{\kern 0.25em\overline{\kern -0.25em \Deltares}{}\xspace}
\def\Lbar{\kern 0.2em\overline{\kern -0.2em\Lambda\kern 0.05em}\kern-0.05em{}\xspace}
\def\Sigbar{\kern 0.2em\overline{\kern -0.2em \Sigma}{}\xspace}
\def\Xibar{\kern 0.2em\overline{\kern -0.2em \Xi}{}\xspace}
\def\Obar{\kern 0.2em\overline{\kern -0.2em \Omega}{}\xspace}
\def\Nbar{\kern 0.2em\overline{\kern -0.2em N}{}\xspace}
\def\Xb{\kern 0.2em\overline{\kern -0.2em X}{}\xspace}

\def\mes        {\mbox{$m_{\rm ES}$}\xspace}

\newcommand{\tev}{\ensuremath{\mathrm{\,Te\kern -0.1em V}}\xspace}
\newcommand{\gev}{\ensuremath{\mathrm{\,Ge\kern -0.1em V}}\xspace}
\newcommand{\mev}{\ensuremath{\mathrm{\,Me\kern -0.1em V}}\xspace}
\newcommand{\kev}{\ensuremath{\mathrm{\,ke\kern -0.1em V}}\xspace}
\newcommand{\ev}{\ensuremath{\mathrm{\,e\kern -0.1em V}}\xspace}
\newcommand{\gevc}{\ensuremath{{\mathrm{\,Ge\kern -0.1em V\!/}c}}\xspace}
\newcommand{\mevc}{\ensuremath{{\mathrm{\,Me\kern -0.1em V\!/}c}}\xspace}
\newcommand{\gevcc}{\ensuremath{{\mathrm{\,Ge\kern -0.1em V\!/}c^2}}\xspace}
\newcommand{\mevcc}{\ensuremath{{\mathrm{\,Me\kern -0.1em V\!/}c^2}}\xspace}
\def\invfb   {\ensuremath{\mbox{\,fb}^{-1}}\xspace}

\def\mus  {\ensuremath{\rm \,\mus}\xspace}

\def\mus        {\ensuremath{\,\mu{\rm s}}\xspace}    %% microsecond
\def\to                 {\ensuremath{\rightarrow}\xspace}

\def\pep2{PEP-II}

\def\gsim{{~\raise.15em\hbox{$>$}\kern-.85em
          \lower.35em\hbox{$\sim$}~}\xspace}
\def\lsim{{~\raise.15em\hbox{$<$}\kern-.85em
          \lower.35em\hbox{$\sim$}~}\xspace}

\def\CP                {\ensuremath{C\!P}\xspace}
\xspace

\newcommand{\jprlBase}       {Phys.\ Rev.\ Lett.\xspace}
\newcommand{\jprBase}        {Phys.\ Rev.\xspace}
\newcommand{\jplBase}        {Phys.\ Lett.\xspace}
\newcommand{\nimBaseA}       {Nucl.\ Instr.\ Methods Phys.\ Res., Sect.\ A\xspace}

\newcommand{\nima}      [1]  {\nimBaseA~{\bf #1}}

\newcommand{\plb}       [1]  {\jplBase\ B~{\bf #1}}

\newcommand{\jprl}      [1]  {\jprlBase\ {\bf #1}}
\newcommand{\jprd}      [1]  {\jprBase\ D~{\bf #1}}

\newcommand{\progtp}    [1]  {{Prog.\ Theor.\ Phys.\ {\bf #1}}}

\def\jetset74   {\mbox{\tt Jetset \hspace{-0.5em}7.\hspace{-0.2em}4}\xspace}
\newcommand{\de}{\ensuremath{\mbox{$\Delta E$}}\xspace}
\newcommand{\fis}{\ensuremath{\mbox{$\mathcal{F}$}}\xspace}

\def\figurebox#1#2#3{%
    \def\arg{#3}%
    \ifx\arg\empty
    {\hfill\vbox{\hsize#2\hrule\hbox to #2{\vrule\hfill\vbox to #1{\hsize#2\vfill}\vrule}\hrule}\hfill}%
    \else
    {\hfill\epsfbox{#3}\hfill}%
    \fi}

\def\Dztilde   {\ensuremath {\tilde{D}^0}\xspace}
\def\Dstarztilde   {\ensuremath {\tilde{D}^{\ast 0}}\xspace}
\def\dodstar {\ensuremath {D^{(\ast)0}}\xspace}
\def\dodstarb {\ensuremath {\Db^{(\ast)0}}\xspace}
\def\dodstartilde {\ensuremath {\tilde{D}^{(\ast)0}}\xspace}
\def\dotokspp {\ensuremath {\Dz\to\KS\pim\pip}\xspace}
\def\dotildetokspp {\ensuremath {\Dztilde\to\KS\pim\pip}\xspace}
\def\dotokspp {\ensuremath {\Dz\to\KS\pim\pip}\xspace}

\def\btodtildek   {\ensuremath {\Bm\to\Dztilde\Km}\xspace}

\def\btodsttildek   {\ensuremath {\Bm\to\Dstarztilde\Km}\xspace}
\def\btoddstk   {\ensuremath {\Bm\to\dodstar\Km}\xspace}
\def\btoddstbk   {\ensuremath {\Bm\to\dodstarb\Km}\xspace}
\def\btoddsttildek   {\ensuremath {\Bm\to\dodstartilde\Km}\xspace}
\def\bmptoddsttildek   {\ensuremath {\Bmp\to\dodstartilde\Kmp}\xspace}

\def \rb {\ensuremath {r_B}\xspace}
\def \rbs {\ensuremath {r^\ast_B}\xspace}
\def \rbbs {\ensuremath {r^{(\ast)}_B}\xspace}
\def \deltab {\ensuremath {\delta_B}\xspace}
\def \deltabs {\ensuremath {\delta^\ast_B}\xspace}
\def \deltabbs {\ensuremath {\delta^{(\ast)}_B}\xspace}

\def \rbbspm {\ensuremath {r^{(\ast)}_{B^\pm}}\xspace}

\def \zbp {\ensuremath {{\mathsf z}_+}\xspace}
\def \zbm {\ensuremath {{\mathsf z}_-}\xspace}

\def \zbsp {\ensuremath {{\mathsf z}_+^\ast}\xspace}
\def \zbsm {\ensuremath {{\mathsf z}_-^\ast}\xspace}

\def \xbbspm {\ensuremath {x_\pm^{(\ast)}}\xspace}
\def \ybbspm {\ensuremath {y_\pm^{(\ast)}}\xspace}
\def \zbbspm {\ensuremath {{\mathsf z}_\pm^{(\ast)}}\xspace}

\def \zbbs {\ensuremath {{\mathsf z}^{(\ast)}}\xspace}

\def\bea{\begin{eqnarray}}
\def\eea{\end{eqnarray}}
\def\nn{\nonumber}

\newcommand{\reslinepp}[6]{#1\pm#2\pm#3\pm#4~[#5,#6]}
\newcommand{\reslineppval}[4]{#1\pm#2\pm#3\pm#4}

\newcommand{\reslinepol}[8]{\left(#1\pm#2~^{+#3}_{-#4}~^{+#5}_{-#6}\right)^\circ~[#7^\circ, #8^\circ]}

\newcommand{\reslinepolval}[6]{\left(#1\pm#2~^{+#3}_{-#4}~^{+#5}_{-#6}\right)^\circ}
\newcommand{\reslinepolvalpp}[5]{\left(#1\pm#2~^{+#3}_{-#4}\pm#5\right)^\circ}

\begin{document}

%\preprint{\babar-PUB-\BABARPubYear/\BABARPubNumber} 
%\preprint{SLAC-PUB-\SLACPubNumber} 

\begin{flushleft}
%\mbox{\normalsize {\babar\ }Analysis Document \#1102, Version 19} \\
%\babar-PUB-\BABARPubYear/\BABARPubNumber ~~~~~~~ SLAC-PUB-\SLACPubNumber   \\
\end{flushleft}

\vskip-2.2cm
\begin{flushright}
%hep-ex/\LANLNumber
\end{flushright} 

\title{
{
\large \bf \boldmath Measurement of the Cabibbo-Kobayashi-Maskawa angle $\gamma$ in $\Bmp \to D^{(*)} \Kmp$ decays with 
a Dalitz analysis of $D \to \KS\pim\pip$
}
}

%author list
%\input pubboard/authors_mar2005.tex
%\input pubboard/authors_pub05010.tex
%% author list as of 02-Mar-2005 (634 authors)
%
\author{B.~Aubert}
\author{R.~Barate}
\author{D.~Boutigny}
\author{F.~Couderc}
\author{Y.~Karyotakis}
\author{J.~P.~Lees}
\author{V.~Poireau}
\author{V.~Tisserand}
\author{A.~Zghiche}
\affiliation{Laboratoire de Physique des Particules, F-74941 Annecy-le-Vieux, France }
\author{E.~Grauges}
\affiliation{IFAE, Universitat Autonoma de Barcelona, E-08193 Bellaterra, Barcelona, Spain }
% ...
\collaboration{The \babar\ Collaboration}
\noaffiliation

\date{\today}% It is always \today, today, but you may specify any date with \date.

\begin{abstract} 
\noindent
We report on a measurement of the Cabibbo-Kobayashi-Maskawa \CP-violating phase $\gamma$ through a Dalitz analysis
of neutral $D$ decays to $\KS\pim\pip$ in the processes $\Bmp \to D^{(*)}\Kmp$,
$\Dstar \to D\piz,D\g$.
Using a sample of 227 million \BB pairs collected by the \babar\ detector, %at the PEP-II %asymmetric-energy \epem\ 
%collider at SLAC, 
we measure the amplitude ratios
%$\rb = \reslinepval{0.118}{0.079}{0.034}{0.036}{0.034}$
$\rb = \reslineppval{0.12}{0.08}{0.03}{0.04}$
and
%$\rbs = \reslineval{0.169}{0.096}{0.030}{0.028}{0.029}{0.026}$,
$\rbs = \reslineppval{0.17}{0.10}{0.03}{0.03}$,
the relative strong phases
$\deltab=\reslinepolval{104}{45}{17}{21}{16}{24}$ 
and 
$\deltabs=\reslinepolvalpp{-64}{41}{14}{12}{15}$ between the amplitudes  
${\cal A} (\btoddstbk)$ and ${\cal A} (\btoddstk)$, and
$\gamma = \reslinepolval{70}{31}{12}{10}{14}{11}$.
The first error is statistical, the second is the experimental
systematic uncertainty and the third reflects the Dalitz model uncertainty. 
The results for the strong and weak phases have a two-fold ambiguity.
\end{abstract}

\maketitle
\begin{figure}[!t]
\begin{center}
\begin{tabular} {rr}  
\includegraphics[height=3.8cm]{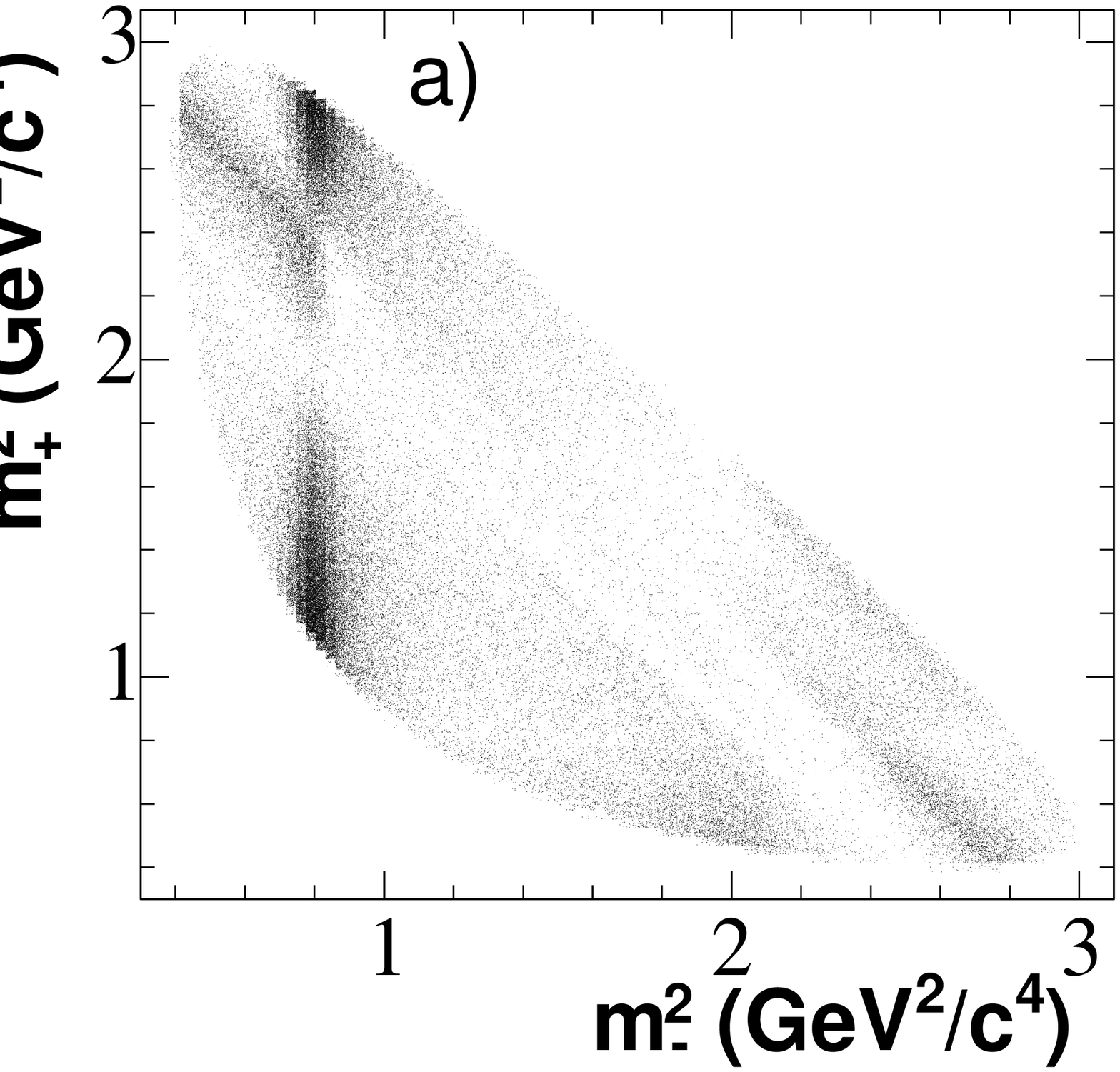} &
\includegraphics[height=3.8cm]{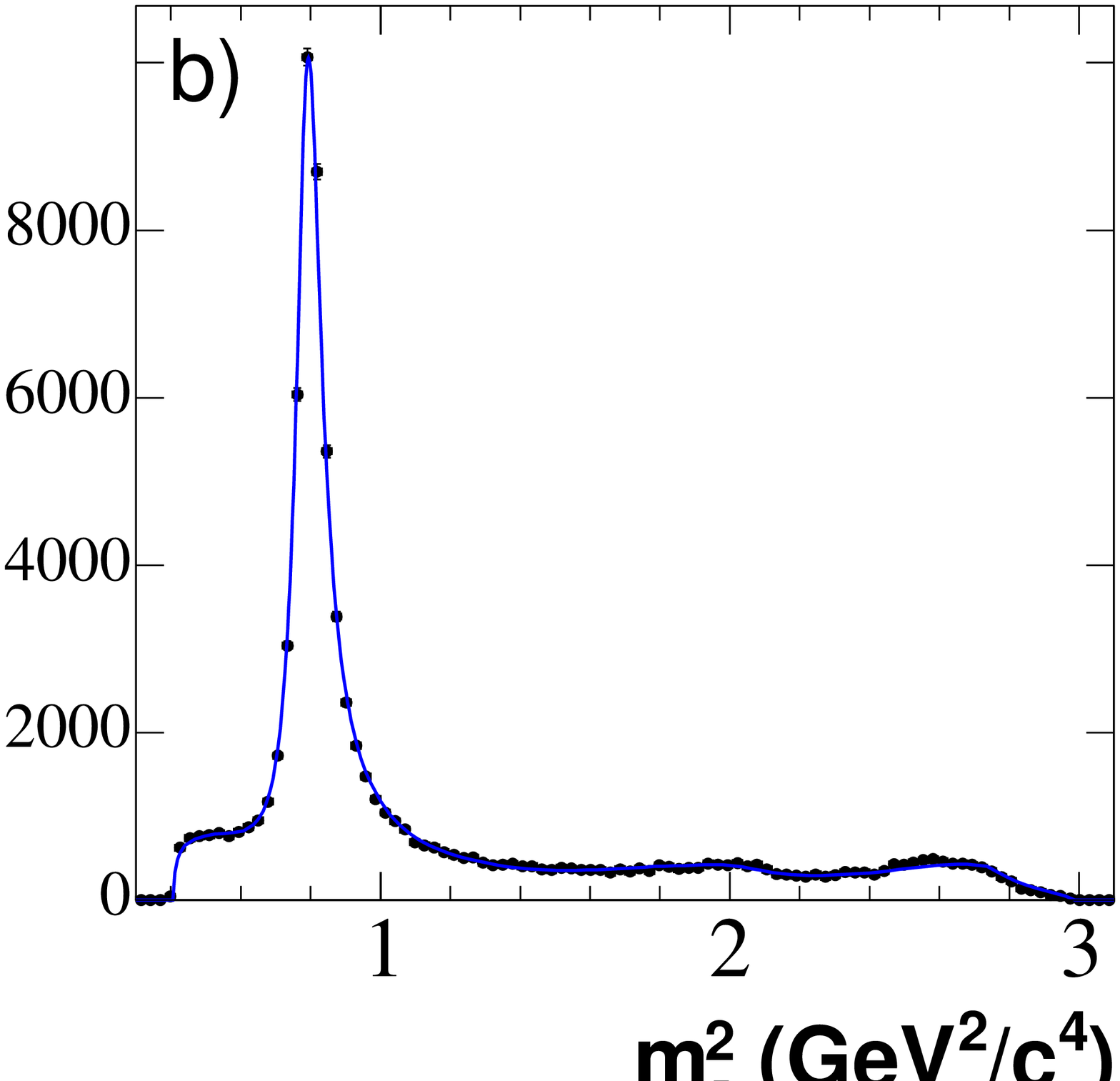} \\
\includegraphics[height=3.8cm]{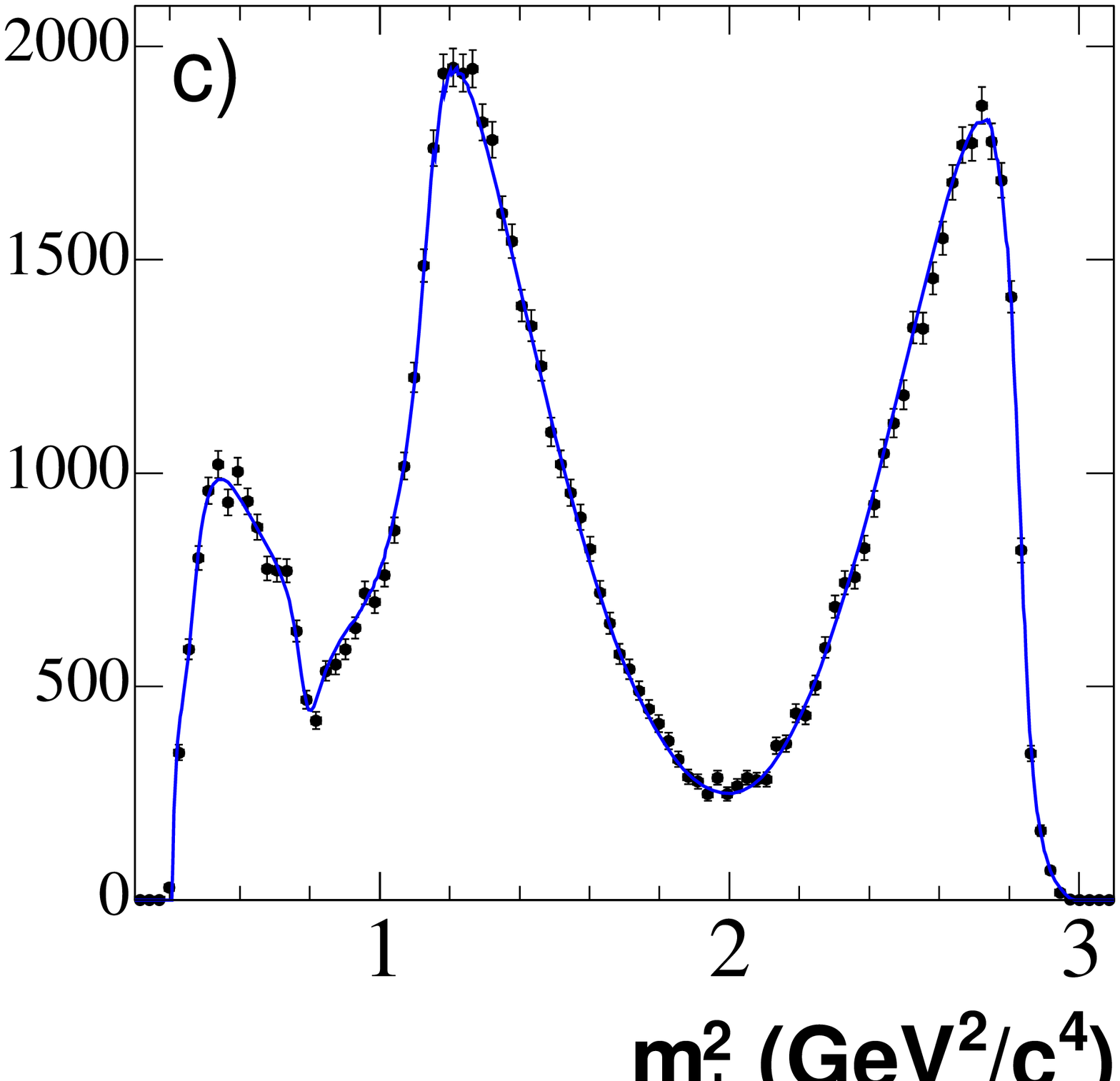} &
\includegraphics[height=3.8cm]{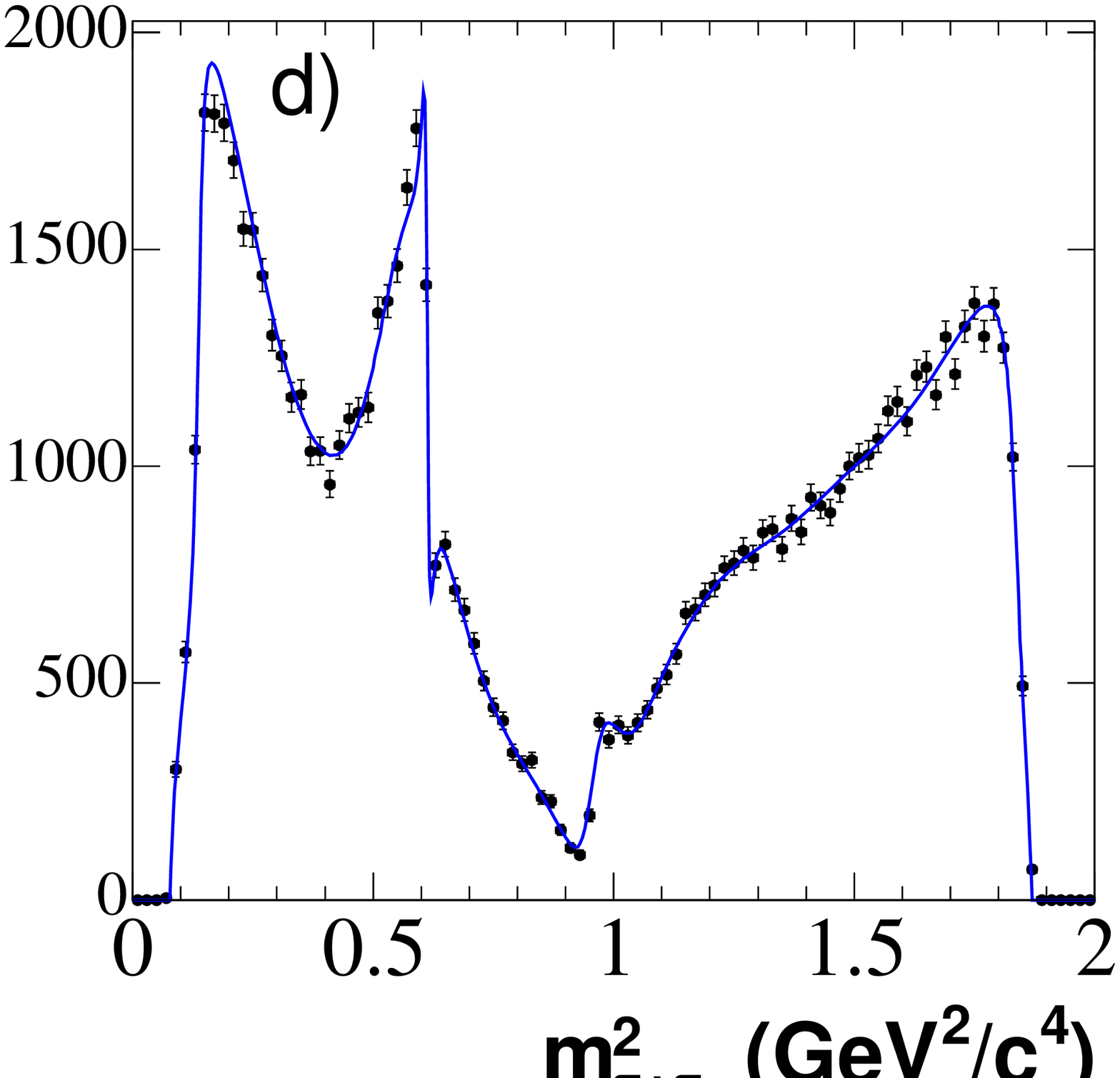} \\
\end{tabular}   
\caption{(a) The $\Dz \to \KS \pim \pip$ Dalitz distribution from $\Dstarp \to \Dz \pip$ events, and projections
on (b) $m^2_-$, (c) $m^2_+$, and (d) $m^2_{\pip\pim}$. The curves are the fit projections.}
\label{fig:dalmkspidcs}
\end{center}
\end{figure}
\begin{figure}[!t]
\begin{center}   
\begin{tabular} {cc} 
 \includegraphics[height=3.9cm]{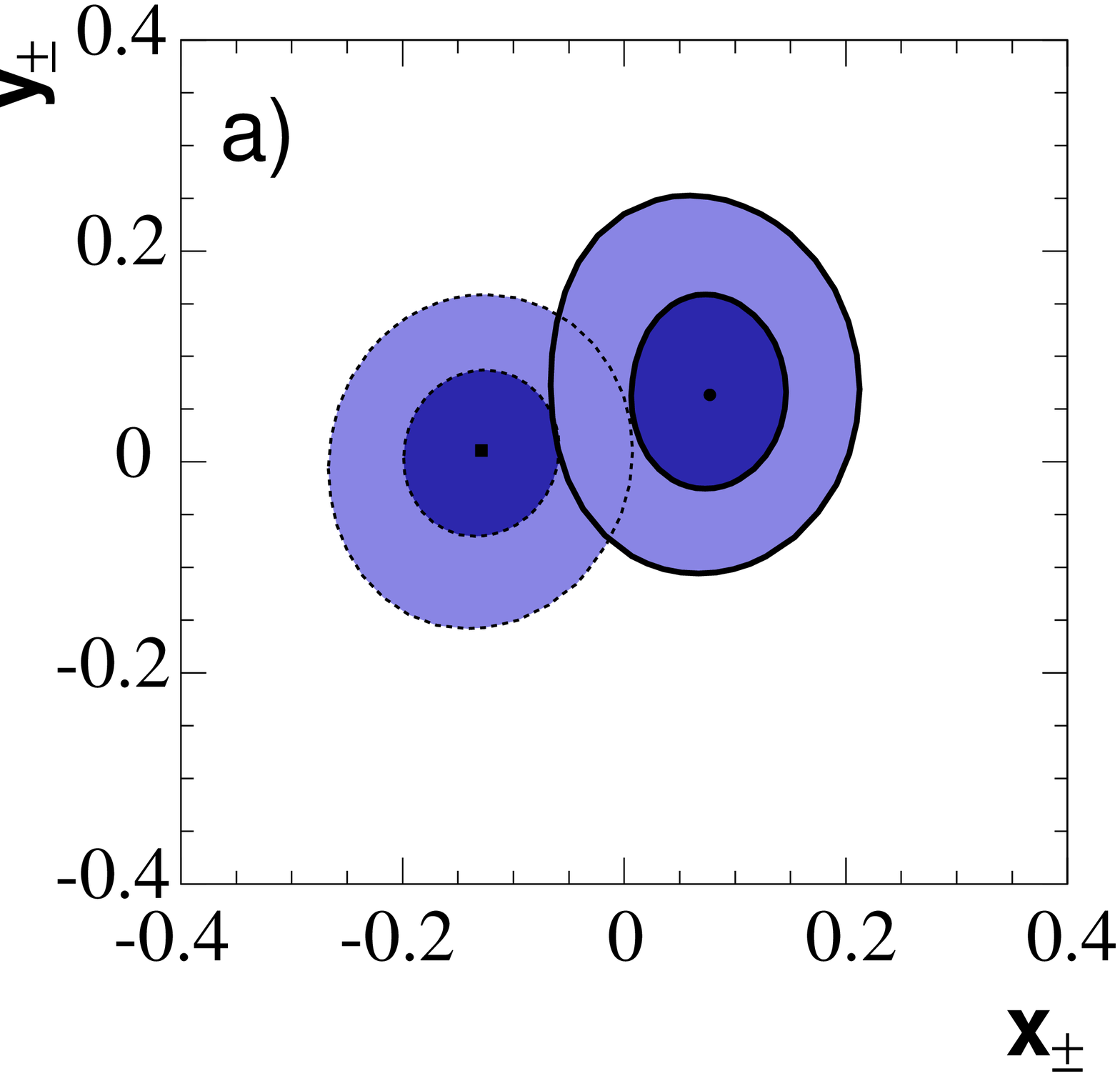} &
 \includegraphics[height=3.9cm]{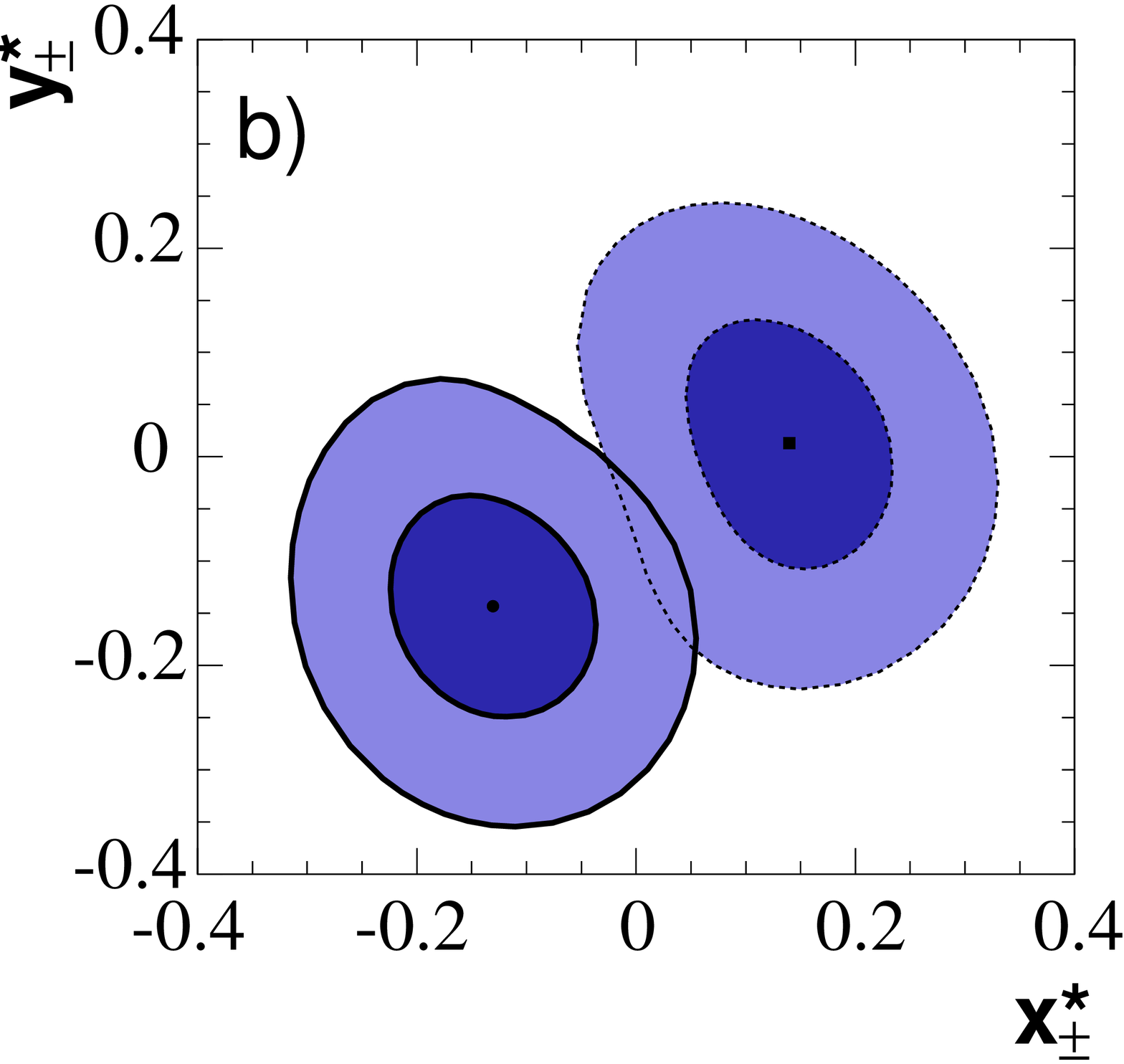} \\
 \includegraphics[height=3.9cm]{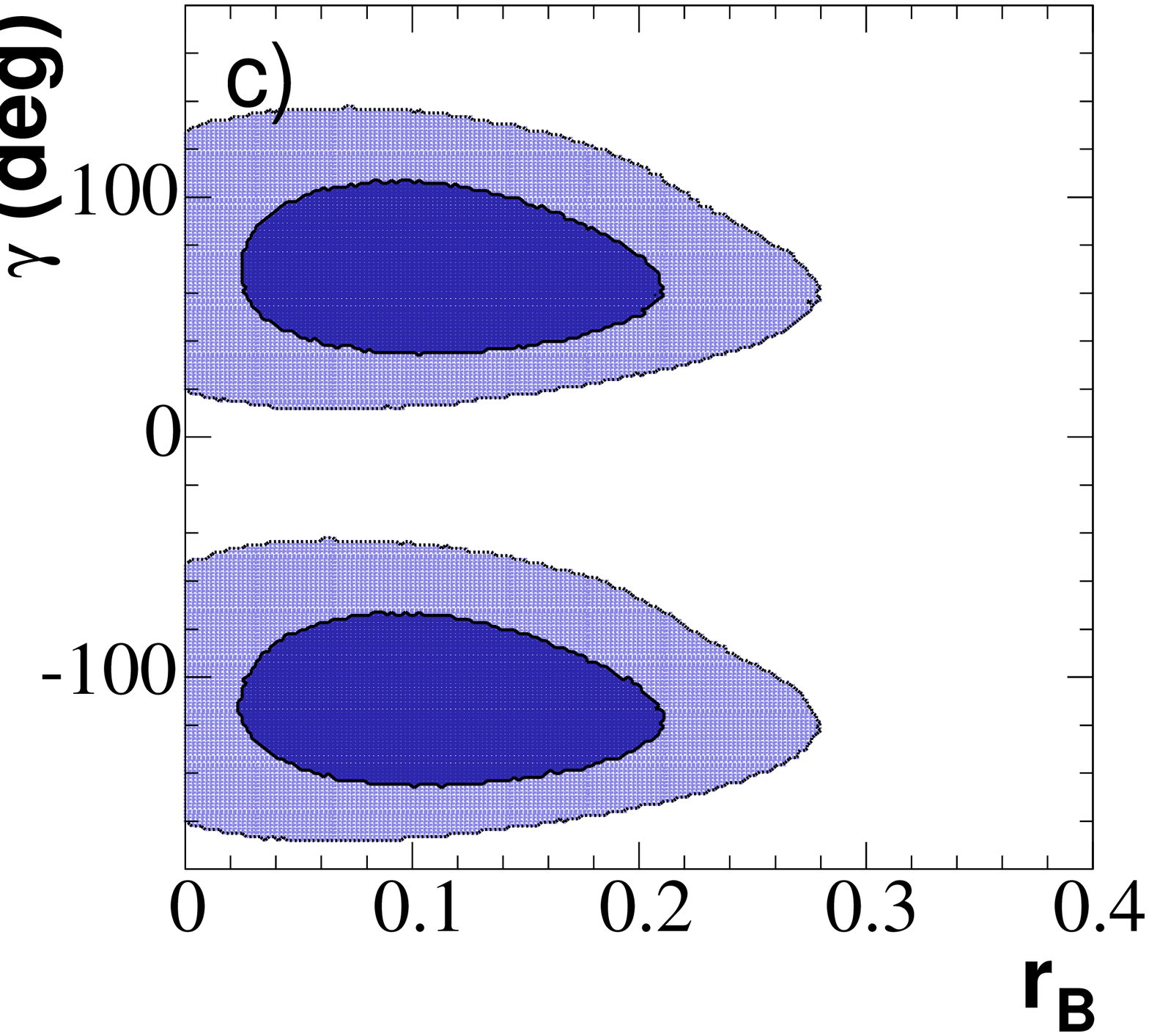} &
 \includegraphics[height=3.9cm]{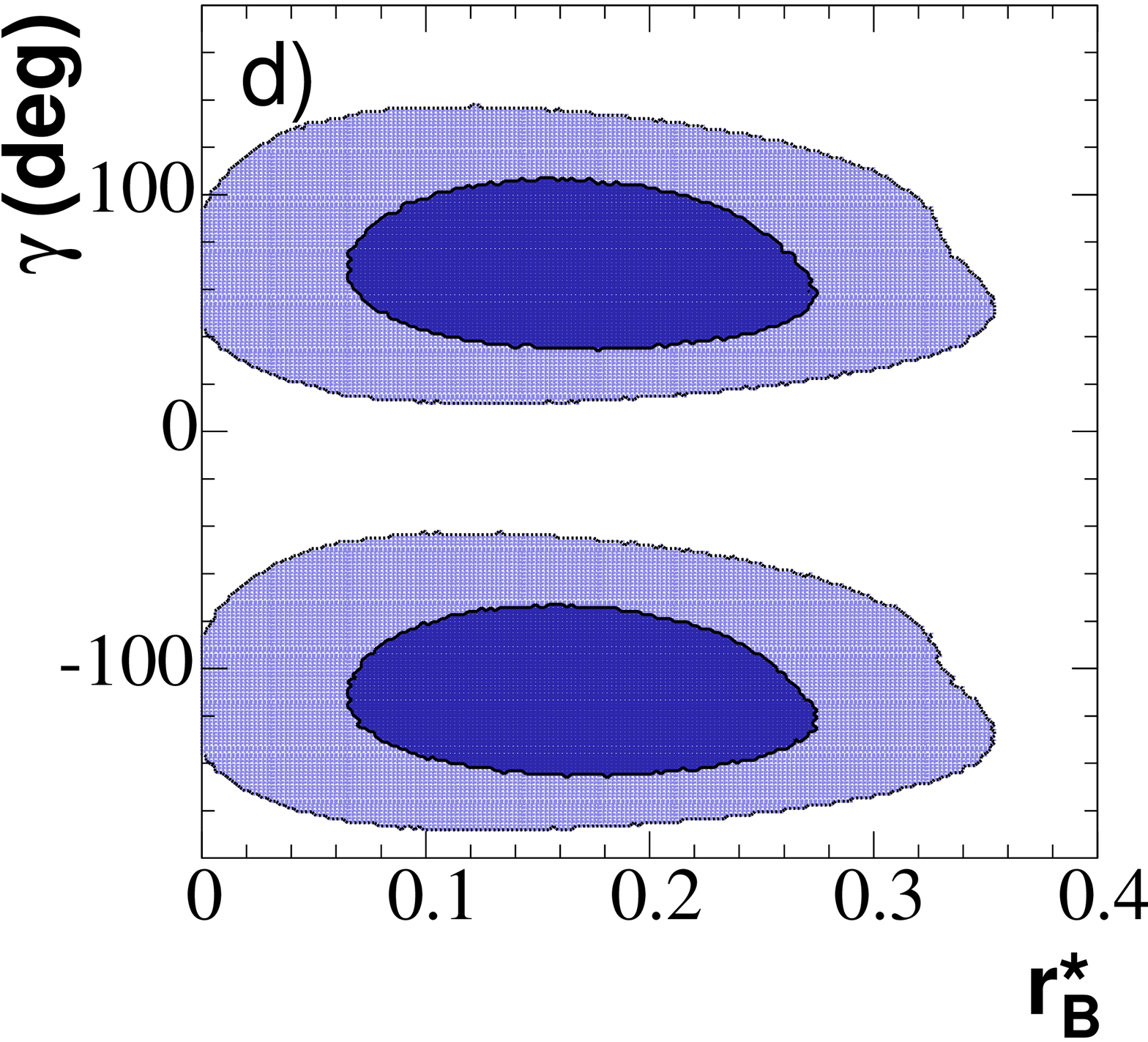}
\end{tabular}   
\caption{
Contours at 39.3\% (dark) and 86.5\% (light) confidence level (statistical only) 
%in the $(\xbbs,\ybbs)$ planes 
in the \zbbs planes
for (a) $\Dztilde\Km$ and (b) $\Dstarztilde\Km$, separately for \Bm (thick and solid lines) 
and \Bp (thin and dotted lines). Projections in the $\rbbs-\g$ planes of the five-dimensional one- (dark) and two- (light)
standard deviation regions, for (c) $\Dztilde\Km$ and (d) $\Dstarztilde\Km$.  
}
\label{fig:contours}
\end{center} 
\end{figure}

\CP violation in the \B meson
system has been clearly established in recent years \cite{ref:CP,ref:CPKpi}. 
%Although there is good agreement with Standard Model expectations, further measurements of 
Although these results are in good agreement with Standard Model expectations, other and
more precise measurements of \CP violation in \B decays are needed to over-constrain the Cabibbo-Kobayashi-Maskawa (CKM)
quark mixing matrix~\cite{ref:CKM} and search for new physics effects.
The angle \g of the unitarity triangle~\cite{ref:pdg2004} of the CKM matrix
constitutes one of these crucial measurements.\\
\indent
Various methods using \btoddsttildek\ \cite{ref:chargeconj} decays have been proposed to 
measure \g~\cite{ref:GLWADS,ref:DKDalitz,ref:bellePRD}. Here, 
\Dztilde indicates either a \Dz or a \Dzb meson and the symbol ``$(*)$'' refers to either a $D$ or \Dstar meson.
All methods exploit the fact that a \Bm can decay 
into a $\dodstar\Km$ ($\dodstarb\Km$) final state via $\b \to \c \ubar \s$ ($\b \to \u \cbar \s$) transitions.
These decay amplitudes interfere when the \Dz and \Dzb decay into the same final
state, which can lead to different \Bp and \Bm decay rates (direct \CP violation). 
%\\
%\indent
In this Letter we report on a measurement of \g based on 
the analysis of the Dalitz distribution of the three-body decay \dotokspp~\cite{ref:DKDalitz,ref:bellePRD}. 
The primary advantage of this method is that it involves the entire resonant structure of the three-body decay, 
%This method involves the entire resonant structure of the three-body decay, 
%with interference of Cabibbo-allowed (CA) and doubly Cabibbo-suppressed (DCS) amplitudes, 
with interference of doubly Cabibbo-suppressed (DCS), Cabibbo-allowed (CA), and \CP eigenstate amplitudes,
providing the sensitivity to \g.
The analysis is based on an integrated luminosity of 205 \invfb recorded at the \FourS resonance
(corresponding to 227 million \BB decays)
and 9.6 \invfb collected at 
a center-of-mass (CM) energy 40 \mev below with the \babar\ detector~\cite{ref:detector} at 
the SLAC \pep2 $e^+e^-$ asymmetric-energy \B Factory.
\\
\indent
The small \CP asymmetry in $D$ decays allowed by the present experimental limits~\cite{ref:DneuCP} 
has a negligible effect on this analysis. Thus, the \bmptoddsttildek, $\Dstarztilde \to \Dztilde\piz,\Dztilde\g$, 
\dotildetokspp decay chain amplitude 
${\cal A}_\mp^{(*)}(m^2_-,m^2_+)$ can be written as  
\bea
{\cal A}_D(m^2_\mp,m^2_\pm) + \kappa\rbbs e^{i(\deltabbs \mp \g)}{\cal A}_D(m^2_\pm,m^2_\mp)~,\nn 
%\label{eq:ampB}
\eea
where $m^2_-$ and $m^2_+$ are the squared invariant masses of the 
$\KS\pim$ and $\KS\pip$ combinations, respectively, and ${\cal A}_D(m^2_-,m^2_+)$ is
the \dotokspp\ decay amplitude.
Here, \rbbs\ and \deltabbs\ are the amplitude ratios and relative strong phases between the amplitudes
\mbox{${\cal A} (\btoddstbk)$} and \mbox{${\cal A} (\btoddstk)$}. As a consequence of parity 
and angular momentum conservation in the $\Dstarztilde$ decay, the factor $\kappa$ takes the value 
$+1$ for \btodtildek and $\Bm\to\Dstarztilde (\Dztilde\piz)\Km$, and $-1$ 
for $\Bm\to\Dstarztilde (\Dztilde\g)\Km$~\cite{ref:bondar}.
We first determine 
${\cal A}_D(m^2_-,m^2_+)$ 
through a Dalitz analysis 
of a high-statistics sample of tagged \Dz mesons from inclusive $\Dstarp \to \Dz \pip$ decays reconstructed in data.
%Having determined 
%${\cal A}_D(m^2_-,m^2_+)$, 
We then perform a simultaneous fit to the 
$|{\cal A}^{(*)}_-(m^2_-,m^2_+)|^2$ and $|{\cal A}^{(*)}_+(m^2_-,m^2_+)|^2$ 
distributions for the \bmptoddsttildek\ samples 
to determine the \CP parameters \rbbs, $\deltabbs$, and \g. 
We emphasize that in this analysis the Dalitz amplitude is only a means to 
extract the \CP parameters.
\\
\indent
\Bm candidates are formed by combining a mass-constrained \dodstar candidate with a 
track identified as a kaon~\cite{ref:detector}.
We accept $\KS\to\pip\pim$ candidates that have a two-pion invariant mass within 9~\mevcc of the \KS mass~\cite{ref:pdg2004} 
and a cosine of the angle between the line connecting the \Dz and \KS decay vertices and the \KS momentum 
(in the plane transverse to the beam) greater than 0.99.
%\KS candidates are reconstructed from pairs of oppositely-charged 
%pions with an invariant mass within 9~\mevcc of the \KS mass~\cite{ref:pdg2004}.
%The two pions are constrained to originate from the same point and the cosine of the angle 
%between the direction transverse to the beam connecting the \Dz and the \KS decay points (transverse flight direction), 
%and the transverse momentum vector is required to be larger than 0.99.
\Dz candidates are selected by requiring the $\KS\pim\pip$ invariant
mass to be within 12~\mevcc of the \Dz mass~\cite{ref:pdg2004}.
The \piz candidates from $\Dstarz\to \Dz\piz$ are formed from pairs of photons with invariant mass 
%in the range $115<m(\g\g)<150$~\mevcc, 
in the range $[115,150]$~\mevcc, 
and with photon energy greater than 30~\mev. 
Photon candidates from $\Dstarz\to \Dz\g$ are selected if their energy is greater than 100~\mev.
%The \Dz candidates are combined with the low energy \piz (\g)
%to form the \Dstarz candidate, with a \Dstarz-\Dz mass difference 
$\Dstarz\to\Dz\piz(\Dz\g)$ candidates are required to have a \Dstarz-\Dz mass difference
within 2.5 (10) \mevcc of its nominal value~\cite{ref:pdg2004}.\\
\begin{figure}[!t]
\begin{center}
\begin{tabular} {ccc}  
\includegraphics[height=4.1cm]{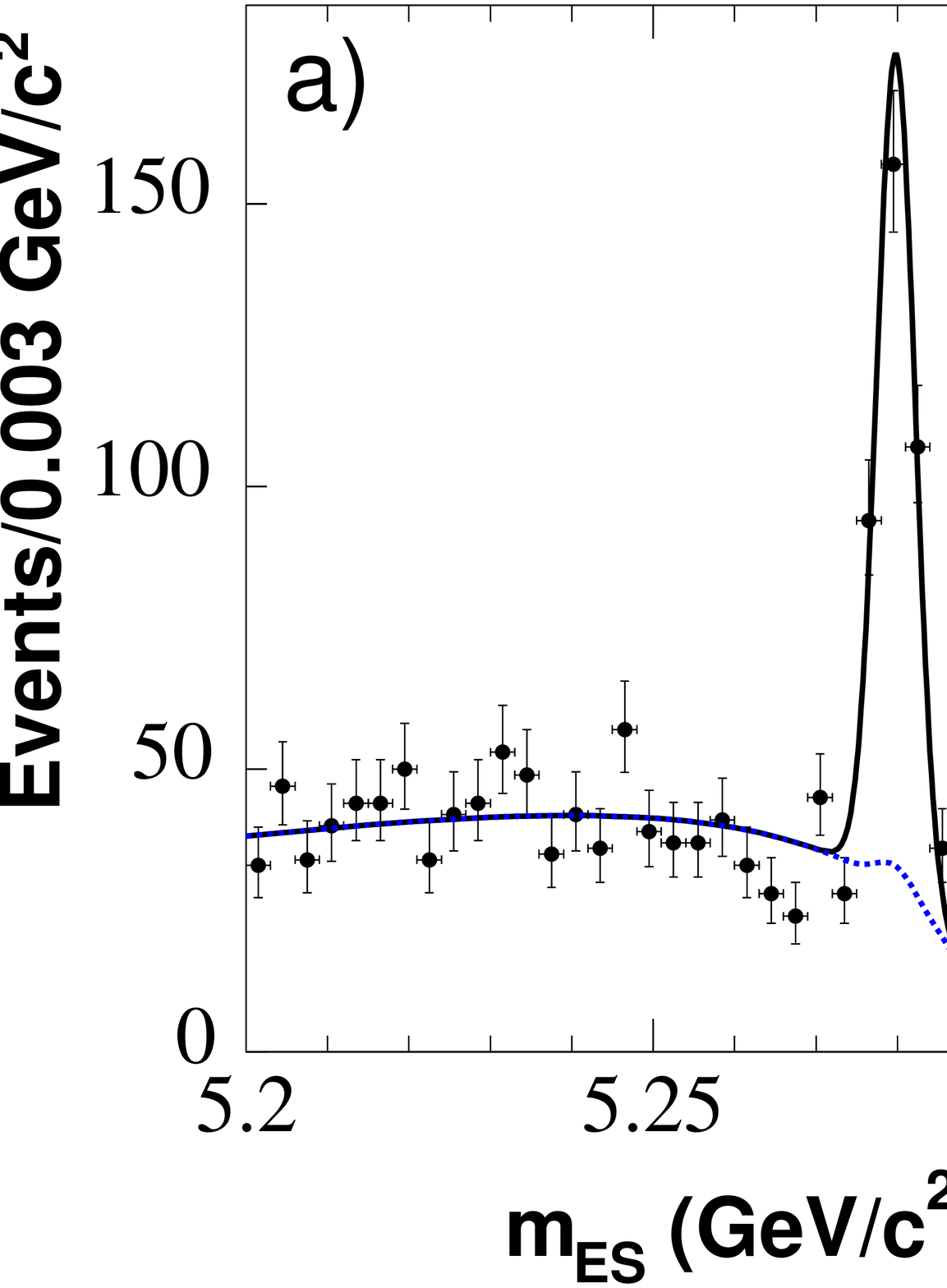} &
\includegraphics[height=4.1cm]{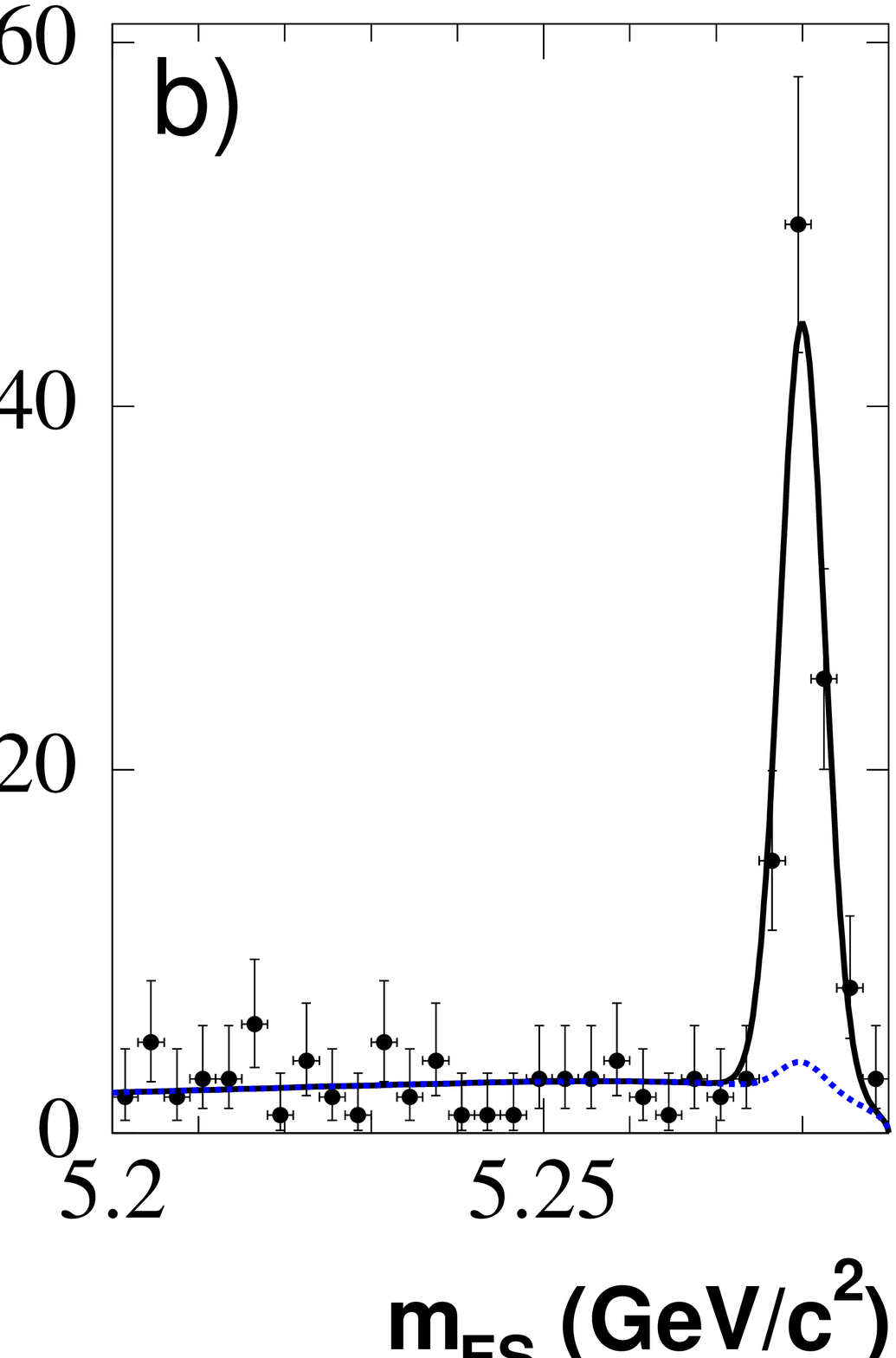} &
\includegraphics[height=4.1cm]{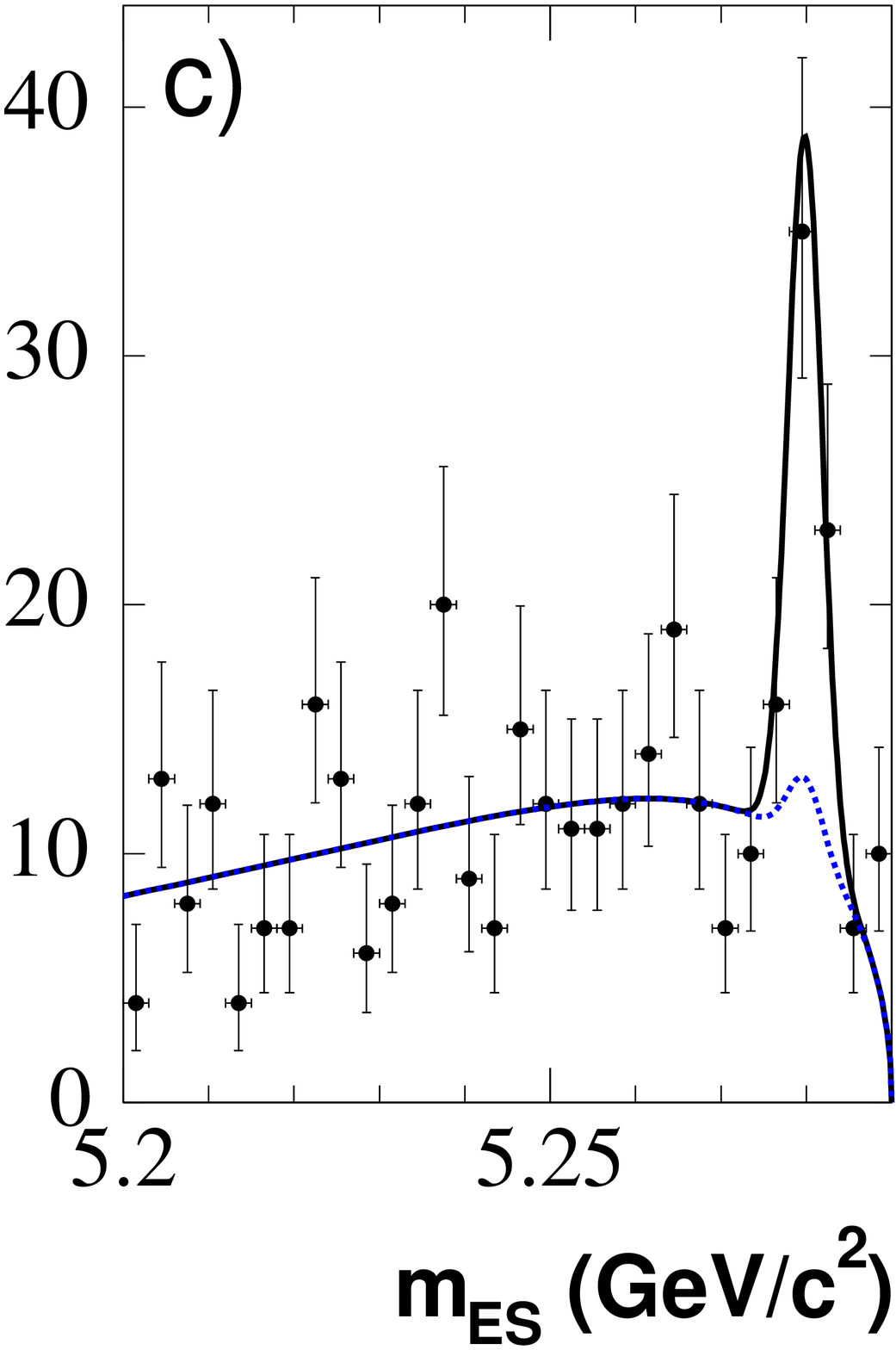}
\end{tabular}   
\caption{Distributions of \mes for (a) \btodtildek, (b) $\Bm\to\Dstarztilde(\Dztilde\piz)\Km$, 
and (c) $\Bm\to\Dstarztilde(\Dztilde\g)\Km$. 
%after all selection criteria are applied. 
The curves represent the fit projections for 
signal plus background (solid lines) and background (dotted lines).
The 
%small 
peaking structure of the background 
%in the signal region 
is due to remaining $\Bm\to\dodstartilde\pim$ events.
}
\label{fig:variables}
\end{center}
\end{figure}
\indent
The beam-energy substituted \B mass 
%\mes~\cite{ref:CPKpi} 
\mes~\cite{ref:kkks-kksks} (Fig.~\ref{fig:variables}) 
and the difference $\de$ between the 
reconstructed energy of the \Bm candidate and the beam energy in the $e^+e^-$ CM frame are used to identify signal \Bm decays.
%The beam-energy substituted mass $\mes=\sqrt{(s/2+{\bf p}_i {\bf p}_\B)^2/E_i^2-p_\B^2}$ and the difference $\de$ between the 
%reconstructed energy of the \Bm candidate in the CM frame and $\sqrt{s}/2$ are used to identify signal \Bm decays. 
%Here, $\sqrt{s}$ is the $e^+e^-$ CM energy, and the \Bm momentum ${\bf p}_\B$ and the four-momentum of the 
%initial $e^+e^-$ state $(E_i,{\bf p}_i)$ are defined in the laboratory frame.
%The \mes distribution for signal events peaks near the \Bm mass with a resolution of $2.4$~\mevcc~(Fig.~\ref{fig:variables}),
%while the \de distribution peaks near zero with a resolution of $15$~\mev. 
We require $\mes>5.2$~\gevcc and $|\de|<30$~\mev.
Since the background is dominated by random combinations of tracks arising 
from $e^+e^- \to \qqbar$, $\q=\{\u,\d,\s,\c\}$ (continuum) events,
we require $|\cos \theta_T^*|<0.8$, where $\theta^*_T$
is the CM angle between the thrust axis of the \Bm candidate and that of the remaining particles in the event.
The reconstruction efficiencies (purities in the signal region $\mes > 5.272$~\gevcc) 
are $18\%$~$(63\%)$, $5.9\%$~$(86\%)$, $8.1\%$~$(52\%)$ for the $\Bm\to\Dztilde\Km$, 
$\Bm\to\Dstarztilde(\Dztilde\piz)\Km$, and $\Bm\to\Dstarztilde(\Dztilde\g)\Km$ decay modes, respectively. 
%For the purpose of determining purities, a signal region $\mes > 5.272$~\gevcc is used for all modes.
The cross-feed among the different samples is negligible.
\\  
\indent
The \Dz\ decay amplitude is determined from an unbinned maximum-likelihood Dalitz fit to
a high-purity (97\%) sample of 81496 $\Dstarp\to\Dz\pip$ decays reconstructed in 91.5 \invfb of data (Fig.~\ref{fig:dalmkspidcs}). 
We use the isobar formalism described in Ref.~\cite{ref:cleomodel} to express 
%${\cal A}_D(m^2_-,m^2_+)$ 
${\cal A}_D$ 
as a sum of two-body decay-matrix elements (subscript $r$) and a non-resonant (subscript NR) contribution,
%
%$${\cal A}_D(m^2_-,m^2_+)  = \Sigma_r a_r e^{i \phi_r} {\cal A}_r(m^2_-,m^2_+) + a_{\rm NR} e^{i \phi_{\rm NR}}~,$$
\bea
{\cal A}_D(m^2_-,m^2_+) = \Sigma_r a_r e^{i \phi_r} {\cal A}_r(m^2_-,m^2_+) + a_{\rm NR} e^{i \phi_{\rm NR}}~,\nn
\eea
where each term is parameterized with an amplitude $a_r$ and a phase $\phi_r$.
The function ${\cal A}_r(m^2_-,m^2_+)$ is the Lorentz-invariant expression for the matrix element of a \Dz meson 
decaying into $\KS \pim \pip$ through an intermediate resonance $r$, parameterized as a
function of the position in the Dalitz plane. 
\\
\indent
Table~\ref{tab:fitreso-likelihood} summarizes the values of $a_r$ and $\phi_r$ obtained 
using a model consisting of 16 two-body elements comprising 13 distinct resonances and 
accounting for efficiency variations across the Dalitz plane and the small background contribution. 
For $r=\rho(770),\rho(1450)$ we use the functional form 
suggested in Ref.~\cite{ref:gounarissakurai}, while the remaining resonances 
are parameterized by a spin-dependent relativistic Breit-Wigner distribution. 
For intermediate states with a \Kstar, the regions of interference between DCS and CA
decays are particularly sensitive to \g, and we include the DCS component when
a significant contribution is expected. In addition, we find that the inclusion of the
scalar $\pi\pi$ resonances $\sigma$ and $\sigma'$ significantly improves the quality 
of the fit~\cite{ref:comment_sigma}. Since the two $\sigma$ resonances are not well established 
and are only introduced to improve the description of our data,
the uncertainty on their existence is considered in the systematic errors.
We estimate the goodness of fit through a two-dimensional $\chi^2$ test 
%using an adaptive binning technique 
and obtain $\chi^2= 3824$ for $3054-32$ degrees of freedom. \\
\begin{table}[!t]
\caption{Amplitudes $a_r$, phases $\phi_r$ and fit fractions 
%of the different components ($\KS\pim$, $\KS\pip$, $\pip\pim$ resonances, and NR) 
obtained from the fit of 
the $\Dz \to \KS\pim\pip$ Dalitz distribution from $\Dstarp \to \Dz \pip$ events. 
Errors are statistical only.
Masses and widths of all resonances except $\sigma$ and $\sigma'$ are taken from~\cite{ref:pdg2004}.  
%The fit fraction is defined as the integral over the Dalitz plane of a single component 
%divided by the coherent sum of all components.  
The fit fraction is defined as the integral of $a_r^2 |{\cal A}_r(m^2_-,m^2_+)|^2$ over the Dalitz plane 
divided by the integral of $|{\cal A}_D(m^2_-,m^2_+)|^2$.
The sum of fit fractions is $1.24$.}
\label{tab:fitreso-likelihood}
\begin{center}
\begin{ruledtabular}
\begin{tabular}{lccc}
    Resonance  &  Amplitude & ~~Phase (deg)  & Fit fraction \\ 
\hline
$K^{*}(892)^-$        &    $  1.781      \pm  0.018   $   &  $ \phantom{-}131.0   \pm  0.8$           &       $0.586$    \\ 
$K^{*}_0(1430)^-$     &    $  2.45       \pm  0.08   $    &  $  \phantom{00}-8.3    \pm  2.5\phantom{..}$   &$0.083$    \\ 
$K^{*}_2(1430)^-$     &    $  1.05       \pm  0.06   $   &  $  \phantom{.}-54.3 \pm  2.6\phantom{.}$   &       $0.027$    \\ 
$K^{*}(1410)^-$       &    $  0.52       \pm  0.09  $    &  $  \phantom{-0.}154 \pm  20\phantom{.}$  &         $0.004$    \\ 
$K^{*}(1680)^-$       &    $  0.89       \pm  0.30  $     &  $  \phantom{.}-139 \pm  14\phantom{.}$   &         $0.003$    \\ 
\hline
$K^{*}(892)^+$     &    $  0.180     \pm  0.008  $   &  $ \phantom{.}-44.1   \pm  2.5\phantom{.}$  &       $0.006$     \\ 
$K^{*}_0(1430)^+$  &    $  0.37      \pm  0.07   $   &  $  \phantom{-00}18 \pm  9\phantom{..}$   &      $0.002$    \\ 
$K^{*}_2(1430)^+$  &    $  0.075     \pm  0.038  $   &  $  \phantom{.}-104 \pm  23\phantom{.}$    &        $0.000$    \\ 
\hline 
$\rho(770)$         &    1 (fixed)                      &    \phantom{00}0 (fixed)                                &        $0.224$    \\ 
$\omega(782)$       &    $  0.0391     \pm  0.0016   $  &  $  \phantom{-}115.3  \pm  2.5$            &        $0.006$     \\ 
$f_0(980) $         &    $  0.482      \pm  0.012\phantom{0}$   &  $-141.8     \pm  2.2$   &        $0.061$     \\ 
$f_0(1370) $        &    $  2.25       \pm  0.30  $    &  $  \phantom{-}113.2     \pm  3.7$          &       $0.032$    \\ 
$f_2(1270) $        &    $  0.922      \pm  0.041  $    &  $  \phantom{0}-21.3  \pm  3.1\phantom{..}$   &     $0.030$     \\ 
$\rho(1450)$        &    $  0.52       \pm  0.09   $   &  $  \phantom{-00.}38 \pm  13\phantom{.}$   &        $0.002$    \\ 
$\sigma$            &    $  1.36       \pm  0.05   $   &  $  -177.9     \pm  2.7  $                &         $0.093$    \\ 
$\sigma'$           &    $  0.340      \pm  0.026   $   &  $ \phantom{-}153.0    \pm  3.8 $               &         $0.013$     \\ 
\hline 
Non Resonant        &    $  3.53       \pm  0.44  $     &  $ \phantom{-}128      \pm  6   $               &         $0.073$  \\ 
\end{tabular}
\end{ruledtabular}
\end{center}
\end{table}
\indent
We simultaneously fit the 
$\Bm\to \dodstartilde \Km$ samples using
an unbinned extended maximum-likelihood fit to extract the \CP-violating parameters along with the signal 
and background yields. Three different background components are considered:
%The selected \Bm sample is composed of $\Bm \to \dodstartilde \Km$ decays (signal), 
continuum events, $\Bm\to\dodstartilde\pim$ and $\FourS \to \BB$ 
(other than $\Bm\to\dodstartilde\pim$) decays. 
%The fit uses \mes, \de, and a Fisher discriminant \fis\ to help to identify continuum background.
%\mes, \de to separate remaining $\Bm\to\dodstartilde\pim$ background remaining after the \de selection cut, 
%and the Fisher discriminant \fis\ described in Ref.~\cite{ref:kkks-kksks} to distinguish signal and continuum background.
In addition to \mes, the fit uses \de and 
a Fisher discriminant~\cite{ref:kkks-kksks} to distinguish signal
from $\Bm\to\dodstartilde\pim$ and continuum background, respectively. 
%\fis\ is a linear combination of four topological variables: $L_0=\sum_{i} p_i^*$, $L_2 =\sum_{i} p_i^*  |\cos \theta^*_i|^2$, 
%and the absolute values of the cosine of the polar angles of the \B candidate
%momentum and thrust direction. Here, $p_i^*$ and $\theta_i^*$ are the
%CM momentum and the angle with respect to
%the \B candidate thrust axis of the remaining tracks and clusters in the event.
%The likelihood for candidate $j$ is obtained by summing the product of the event yield $N_c$, 
%the probability density functions (PDF's) for the kinematic and event shape variables ${\cal P}_{c}$,
%and  the Dalitz distributions ${\cal P}_{c}^{\rm Dalitz}$, over the
%signal and background components $c$. 
%The total log-likelihood 
The log-likelihood
%for each \Bm
%$\Bm\to\Dztilde\Km$, $\Bm\to\Dstarztilde(\Dztilde\piz)\Km$, and $\Bm\to\Dstarztilde(\Dztilde\g)\Km$ 
%sample 
is
\bea
%{\cal L} = \exp\left(-\sum_c N_c\right) \prod_j \sum_c N_c{\cal P}_c(\vec{\xi}_j){\cal P}^{\rm Dalitz}_c(\vec{\eta}_j)~,\nn
\ln {\cal L} = -\sum_c N_c + \sum_j \ln \left[ \sum_c N_c{\cal P}_c(\vec{\xi}_j){\cal P}^{\rm Dalitz}_c(\vec{\eta}_j) \right]~,\nn
\eea
where $\vec{\xi}_j = \{\mes,\de,\fis\}_j$ and $\vec{\eta}_j = (m_-^2,m_+^2)_j$
characterize the event $j$.
Here, ${\cal P}_{c}(\vec{\xi})$ and ${\cal P}_{c}^{\rm Dalitz}(\vec{\eta})$ are the probability density functions (PDF's),
and $N_c$ the event yield for signal or background component $c$.
For signal events, ${\cal P}^{\rm Dalitz}_c(\vec{\eta})$ is given by $|{\cal A}_\mp^{(*)}(\vec{\eta})|^2$ 
corrected by the efficiency variations.
%For signal events, ${\cal P}_{\rm sig,\mp}^{\rm Dalitz}(\vec{\eta}) = \epsilon(\vec{\eta}) {\cal A}_\mp(\vec{\eta})$, where
%$\epsilon(\vec{\eta})$ parameterizes the efficiency variations. 
%and ${\cal A}(\vec{\eta})$ is given by Eq.~(\ref{eq:ampB}).
%The parameters describing the \mes, \de and Dalitz PDF's for \BB background events are determined 
%from a detailed Monte Carlo simulation.  
All PDF shape parameters used to describe signal, 
continuum and $\Bm\to\dodstartilde\pim$ components
are determined directly from $\Bm\to\dodstartilde\Km$ and $\Bm\to\dodstartilde\pim$ signal, sideband regions, and off-peak data,
and are fixed in the final fit for \CP\ parameters and event yields.
Only the \mes, \de and Dalitz PDF's for \BB background events are determined 
from a detailed Monte Carlo simulation.
$\Bm\to\dodstartilde\pim$ candidates have been selected using criteria similar to those applied for  
$\Bm\to\dodstartilde\Km$ but requiring the bachelor pion not to be consistent with the kaon hypothesis.\\
\indent
The \CP fit yields $282\pm20$, $90\pm11$, and $44\pm8$ signal $\Dztilde\Km$, $\Dstarztilde(\Dztilde\piz)\Km$, 
and $\Dstarztilde(\Dztilde\g)\Km$
candidates, respectively, consistent with expectations based on measured branching fractions 
and efficiencies estimated from Monte Carlo simulation.
The results for the \CP-violating parameters 
$\zbbspm \equiv (\xbbspm,\ybbspm)$, where $\xbbspm$ and $\ybbspm$ 
are defined as the real and imaginary parts
of the complex amplitude ratios $\rbbspm e^{i(\deltabbs \pm \g)}$, respectively, are summarized in Table~\ref{tab:xyresults}.
Here, $\rbbspm$ is the amplitude ratio between the amplitudes $\b\to\u$ and $\b\to\c$, separately 
for \Bp and \Bm.
The only non-zero statistical correlations involving the \CP parameters are 
%for the pairs $(\xbm,\ybm)$, $(\xbp,\ybp)$, $(\xbsm,\ybsm)$, and $(\xbsp,\ybsp)$, 
for the pairs \zbm, \zbp, \zbsm, and \zbsp,
which amount to 
$3\%$, $6\%$, $-17\%$, and $-27\%$, respectively.
%The $\xbbspm$ and $\ybbspm$ variables 
The \zbbspm variables
are more suitable fit parameters than 
\rbbs, \deltabbs and \g because they are better behaved near the origin, especially in low-statistics samples.
Figures~\ref{fig:contours}(a,b) show the 
one- and two-standard deviation confidence-level contours (statistical only)
%in the $(\xbbs,\ybbs)$ planes 
in the \zbbs planes
for $\Dztilde\Km$ and $\Dstarztilde\Km$, and separately 
for \Bm and \Bp. The separation between 
the \Bm and \Bp regions 
in these planes is an indication of direct \CP violation.\\
%
%\begin{table}[!t]
%\caption{\CP-violating parameters $(\xbbspm,\ybbspm)$ obtained from the \CP fit to the $\Bm\to\dodstartilde\Km$ samples.
%The first error is statistical, the second is the experimental
%systematic uncertainty and the third reflects the Dalitz model uncertainty. 
%}
%\label{tab:xyresults}
%\begin{center}
%\begin{ruledtabular}
%\begin{tabular}{lc}
%\CP parameter & Result \\ \hline
%$\xbm \equiv  \re \{\rbm e^{i(\deltab-\g)}\}$ & $\phantom{-}0.077\pm0.069\pm0.026\pm0.019$ \\
%$\ybm \equiv  \im \{\rbm e^{i(\deltab-\g)}\}$ & $\phantom{-}0.064\pm0.092\pm0.037\pm0.042$ \\ 
%$\xbp \equiv  \re \{\rbp e^{i(\deltab+\g)}\}$ & $-0.129\pm0.070\pm0.030\pm0.032$ \\ 
%$\ybp \equiv  \im \{\rbp e^{i(\deltab+\g)}\}$ & $\phantom{-}0.019\pm0.079\pm0.023\pm0.021$ \\ 
%$\xbsm \equiv  \re \{\rbsm e^{i(\deltabs-\g)}\}$ & $-0.131\pm0.093\pm0.028\pm0.021$ \\
%$\ybsm \equiv  \im \{\rbsm e^{i(\deltabs-\g)}\}$ & $-0.143\pm0.105\pm0.022\pm0.025$ \\
%$\xbsp \equiv  \re \{\rbsp e^{i(\deltabs+\g)}\}$ & $\phantom{-}0.140\pm0.093\pm0.028\pm0.025$ \\
%$\ybsp \equiv  \im \{\rbsp e^{i(\deltabs+\g)}\}$ & $\phantom{-}0.013\pm0.120\pm0.037\pm0.056$ \\
%\end{tabular}
%\end{ruledtabular}
%\end{center}
%\end{table}
%
%
\begin{table}[!t]
\caption{\CP-violating 
%parameters $(\xbbspm,\ybbspm)$ obtained 
parameters \zbbspm obtained 
from the \CP fit to the $\Bm\to\dodstartilde\Km$ samples.
The first error is statistical, the second is the experimental
systematic uncertainty and the third reflects the Dalitz model uncertainty.
}
\label{tab:xyresults}
\begin{center}
\begin{ruledtabular}
\begin{tabular}{lcc}
           & \xbbspm & \ybbspm \\ [0.025in] \hline 
 $\zbm$    & $\phantom{-}0.08\pm0.07\pm0.03\pm0.02$ & $\phantom{-}0.06\pm0.09\pm0.04\pm0.04$ \\
 $\zbp$    & $-0.13\pm0.07\pm0.03\pm0.03$ & $\phantom{-}0.02\pm0.08\pm0.02\pm0.02$ \\
 $\zbsm$   & $-0.13\pm0.09\pm0.03\pm0.02$ & $-0.14\pm0.11\pm0.02\pm0.03$ \\
 $\zbsp$   & $\phantom{-}0.14\pm0.09\pm0.03\pm0.03$ & $\phantom{-}0.01\pm0.12\pm0.04\pm0.06$ \\
\end{tabular}
\end{ruledtabular}
\end{center}
\end{table}
\indent
The largest single contribution to the systematic uncertainties in the \CP parameters 
comes from the choice of the Dalitz model used to describe 
the $\Dz\to\KS\pim\pip$ decay amplitudes. To evaluate this uncertainty
we use the nominal Dalitz model (Table~\ref{tab:fitreso-likelihood}) to generate
large samples of pseudo-experiments. 
We then compare experiment by experiment the 
%values of \xbbspm and \ybbspm obtained from
values of \zbbspm obtained from
fits using the nominal model and a set of alternative models.
We find that removing different combinations of \Kstar and $\rho$ resonances 
(with low fit fractions), 
or changing the functional form of the resonance shapes, has little effect on the total $\chi^2$ of the
fit, or on the 
%values of \xbbspm and \ybbspm.
values of \zbbspm.
However, models where
one or both of the $\sigma$ resonances are removed lead to a
significant increase in the $\chi^2$ of the fit. We use the average 
%variations of \xbbspm and \ybbspm corresponding 
variations of \zbbspm corresponding
to this second set of alternative models as the systematic 
uncertainty due to imperfect knowledge of ${\cal A}_D$.\\
\indent
The experimental systematic uncertainties include the errors on the \mes, \de, and \fis\ PDF parameters for 
signal and background, the uncertainties in the
knowledge of the Dalitz distribution of background events, the efficiency variations across the Dalitz plane, 
and the uncertainty in the fraction of events with a
real \Dz produced in a back-to-back configuration with a negatively-charged kaon.
Less significant systematic uncertainties originate from the imprecise knowledge of the fraction of real \Dz's,
the invariant mass resolution, and the statistical errors in the Dalitz amplitudes and phases from
the fit to the tagged \Dz sample.
The possible effect of \CP violation in $\Bm\to\dodstartilde\pim$ decays and \BB background
was found to be negligible.\\
\indent
%A frequentist analysis~\cite{ref:pdg2004} has been used to interpret the constraints on \xbbspm and \ybbspm
%in terms of ${\bf p} \equiv (\rb,\rbs,\deltab,\deltabs,\g)$. 
A frequentist (Neyman) construction of the confidence regions of ${\bf p} \equiv (\rb,\rbs,\deltab,\deltabs,\g)$
based on the 
%constraints on \xbbspm and \ybbspm has been 
constraints on \zbbspm has been
adopted~\cite{ref:pdg2004}.
%Based on
Using 
a large number of pseudo-experiments
corresponding to the nominal \CP fit model but with many different values of the \CP fit parameters, we
construct an analytical (Gaussian) parameterization of 
%the PDF of $(\xbbspm,\ybbspm)$ as a function 
the PDF of \zbbspm as a function
of ${\bf p}$. 
For a given ${\bf p}$, the five-dimensional confidence level ${\cal C}=1-\alpha$ is calculated 
%analytically 
by integrating over all points in the fit
parameter space closer (larger PDF) to ${\bf p}$ than the fitted data values. 
The one- (two-) standard deviation region of the \CP parameters is defined as the set of ${\bf p}$ values for
which $\alpha$ is smaller than 3.7\% (45.1\%).\\
%(42.8\%) (45.1\%)
%
%
%\begin{figure}[!t]
%\begin{center}   
%\begin{tabular} {cc} 
% \includegraphics[height=3.9cm]{rbDzK-gamma-DK.eps} &
% \includegraphics[height=3.9cm]{rbDstK-gamma-DK.eps}
%\end{tabular}   
%\caption{Two-dimensional projections in the $\rbbs-\g$ planes of the five-dimensional one- (dark) and two- (light)
%standard deviation regions, for (a) $\Dztilde\Km$ and (b) $\Dstarztilde\Km$.}
%\label{fig:contours-rb-gamma}
%\end{center} 
%\end{figure}
%
%
\indent
Figures~\ref{fig:contours}(c,d) show the two-dimensional projections in 
the $\rbbs-\g$ planes, including systematic uncertainties, for $\Dztilde\Km$ and $\Dstarztilde\Km$. 
The figures show that this Dalitz analysis has a two-fold ambiguity,
$(\g,\deltabbs) \to (\g + 180^\circ, \deltabbs + 180^\circ)$.
The significance of direct \CP violation, obtained by evaluating
${\cal C}$ for the most probable \CP conserving point, corresponds to 1.6, 2.1, and 2.4 standard deviations, 
for $\Dztilde\Km$ and $\Dstarztilde\Km$, and their combination, respectively.
Similar results are obtained using a Bayesian technique with uniform a priori probability distributions
for \rbbs, \deltabbs and \g.\\
\indent
In summary, we have measured the direct \CP-violating parameters in \btoddsttildek\ using
a Dalitz analysis of \dotildetokspp\ decays, obtaining 
$\rb = \reslinepp{0.12}{0.08}{0.03}{0.04}{0}{0.28}$,
$\rbs = \reslinepp{0.17}{0.10}{0.03}{0.03}{0}{0.35}$,
%$\rb = \reslinep{0.118}{0.079}{0.034}{0.036}{0.034}{0}{0.279}$,
%$\rbs = \resline{0.169}{0.096}{0.030}{0.028}{0.029}{0.026}{0}{0.354}$,
$\deltab=\reslinepolval{104}{45}{17}{21}{16}{24}$, 
$\deltabs=\reslinepolvalpp{-64}{41}{14}{12}{15}$,
and
$\gamma = \reslinepol{70}{31}{12}{10}{14}{11}{12}{137}$. 
The first error is statistical, the second is the experimental
systematic uncertainty and the third reflects the Dalitz model uncertainty. 
The values inside square brackets indicate the two-standard deviation intervals.
The results for $\gamma$ from \btodtildek and \btodsttildek alone are $(70\pm38)^\circ$ and $(71\pm35)^\circ$, 
respectively (statistical errors only). The constraint on %the weak phase 
\g is consistent with that reported by the Belle Collaboration~\cite{ref:bellePRD}, which has a slightly better statistical precision 
since our \rbbs constraint favors smaller values.
%The results are consistent with and have a similar precision to those reported by the Belle Collaboration~\cite{ref:bellePRD}.
%\\
\indent
We are grateful for the excellent luminosity and machine conditions
provided by our \pep2\ colleagues, 
and for the substantial dedicated effort from
the computing organizations that support \babar.
The collaborating institutions wish to thank 
SLAC for its support and kind hospitality. 
This work is supported by
DOE
and NSF (USA),
NSERC (Canada),
IHEP (China),
CEA and
CNRS-IN2P3
(France),
BMBF and DFG
(Germany),
INFN (Italy),
FOM (The Netherlands),
NFR (Norway),
MIST (Russia), and
PPARC (United Kingdom). 
Individuals have received support from CONACyT (Mexico), A.~P.~Sloan Foundation, 
Research Corporation,
and Alexander von Humboldt Foundation.


\begin{thebibliography}{99}
\bibitem{ref:chargeconj} %The symbol \Dztilde indicates either a \Dz or a \Dzb. 
                     %Similarly, the symbol ``$(*)$'' refers either to a neutral $D$ or \Dstar meson.
                     Reference to the charge-conjugate state is implied here and throughout the text unless otherwise stated.
\bibitem{ref:comment_sigma} The $\sigma$ and $\sigma'$ masses and widths are determined %directly 
from the data. We find (in~\mevcc) $M_{\sigma}=484\pm9$, $\Gamma_{\sigma}=383\pm14$, 
$M_{\sigma'}=1014\pm7$, and $\Gamma_{\sigma'}=88\pm13$. 
Errors are statistical.% only.
\end{thebibliography}
\end{document}